\documentclass[pra,showpacs,twocolumn,aps]{revtex4}%
\usepackage{graphicx}
\usepackage{epsfig}
\usepackage{amsmath}
\usepackage{amsfonts}
\usepackage{amssymb}%
\newcommand{\be}{\begin{equation}}
\newcommand{\ee}{\end{equation}}
\newcommand{\bel}[1]{\begin{equation}\label{#1}}
\newcommand{\bea}{\begin{eqnarray}}
\newcommand{\eea}{\end{eqnarray}}
\newcommand{\ba}{\begin{array}}
\newcommand{\ea}{\end{array}}
\def\bbbc{{\mathchoice {\setbox0=\hbox{$\displaystyle\rm C$}\hbox{\hbox
to0pt{\kern0.4\wd0\vrule height0.9\ht0\hss}\box0}}
{\setbox0=\hbox{$\textstyle\rm C$}\hbox{\hbox
to0pt{\kern0.4\wd0\vrule height0.9\ht0\hss}\box0}}
{\setbox0=\hbox{$\scriptstyle\rm C$}\hbox{\hbox
to0pt{\kern0.4\wd0\vrule height0.9\ht0\hss}\box0}}
{\setbox0=\hbox{$\scriptscriptstyle\rm C$}\hbox{\hbox
to0pt{\kern0.4\wd0\vrule height0.9\ht0\hss}\box0}}}}
\begin{document}
\title{Entangling power of permutation invariant quantum states}
\author{Vladislav Popkov}
\affiliation{Institut f\"{u}r Theoretische Physik, Universit\"{a}t zu K\"{o}ln,
Z\"{u}lpicher Str. 77, 50937 K\"{o}ln, Germany.}
\author{Mario Salerno}
\affiliation{Dipartimento di Fisica \textquotedblleft E.R. Caianiello", }
\affiliation{Istituto Nazionale di Fisica Nucleare (INFN) Sezione di Napoli-Gruppo
Collegato di Salerno, Consorzio Nazionale Interuniversitario per le Scienze
Fisiche della Materia (CNISM), Universit\'{a} di Salerno, I-84081 Baronissi
(SA), Italy}
\author{Gunter Sch\"{u}tz}
\affiliation{Institut f\"{u}r Festk\"{o}rperforschung, Forschungszentrum J\"{u}lich - 52425
J\"{u}lich, Germany}
\thanks{E-mail: popkov@thp.uni-koeln.de}
\thanks{E-mail: salerno@sa.infn.it}
\thanks{E-mail: g.schuetz@fz-juelich.de}

\begin{abstract}
We investigate the von Neumann entanglement entropy as function of the size of
a subsystem for permutation invariant ground states in models with finite
number of states per site, e.g., in quantum spin models. We demonstrate that
the entanglement entropy of $n$ sites in a system of length $L$ generically
grows as $\sigma\log_{2}[2\pi en(L-n)/L]+C$, where $\sigma$ is the on-site
spin and $C$ is a function depending only on magnetization.

\end{abstract}
\date{\today }

\pacs{03.67.-a,03.65.Ud,03.67.Hk,75.10.Jm}
\maketitle

\section{Introduction}

Recently it has been argued that for critical (gapless) quantum spin systems
the entanglement entropy of a block of $n$ spins diverges logarithmically as
$\gamma\log_{2}n$, while for non-critical systems it converges to a constant
finite value \cite{Fazio,Vidal,Latorre_condmat}. This property was interpreted
in the framework of conformal field theory \cite{Korepin} associated with the
corresponding quantum phase transition and the prefactor $\gamma$ of the
logarithm was related to the central charge $c$ of the theory as $c=3\gamma$.
This was shown explicitly for the exactly solvable antiferromagnetic
Heisenberg spin $1/2$ chain, i.e., the $XXZ$ model where the anisotropy
parameter $\Delta$ belongs in the critical regime to the interval $(-1,1)$ and
$\gamma=1/3$. Rather surprisingly, at the transition point $\Delta=-1$ from
gapless to noncritical behaviour, the entanglement (von Neumann) entropy of a
block of spins in the ground state was found to grow \textit{faster} than in
the critical domain $-1<\Delta\leq1$, namely as $\gamma\log_{2}n$ with the
logarithmic prefactor satisfying $\frac{1}{2}\leq\gamma\leq1$
\cite{Entanglement_Heisenberg}. At $\Delta=-1$ the ground state of the $XXZ$
Hamiltonian has permutational invariance (up to a gauge transformation), and
is degenerate with respect to the total $z$-magnetization, so that the whole
system is generically described via a density matrix which can be
written as a (weighted) sum of projectors on the
multiplet components. The lower bound $\gamma=\frac{1}{2}$
is attained if the state of the whole system is a \textit{pure state} (only
one component of the multiplet is present), while the upper bound $\gamma=1$
is reached for a \textit{mixed state} where all the components of the
multiplet have equal weights. Note that the pure state is generically not
factorizable, see (\ref{GS}) below, thus producing a mixed state after 
partial tracing.

In the present paper we generalize the approach of
\cite{Entanglement_Heisenberg} to arbitrary permutation invariant quantum spin
states. In particular, we consider the case of a ferromagnetic spin chain with
arbitrary spin $\sigma$ on every site and show that the entanglement entropy
of a block of $n$ spins in the ferromagnetic ground state generically
diverges as $\sigma\log_{2}n$. Our approach uses the invariance of the ground
state under site permutations, thus allowing us to compute the entanglement
entropy exactly for blocks of arbitrary size and for systems of arbitrary length.

The paper is organized as follows. In section 2 we introduce the permutation
invariant states and list physical systems whose ground states have this
symmetry. In section 3 we formulate a theorem which gives the analytical
expression of the eigenvalues of the reduced density matrix for arbitrary spin
$\sigma$. Using this theorem we compute the entanglement entropy of a block of
size $n$ in the finite system of total length $L$ in specific ground state
sectors. Taking the limit of large subsystem sizes, we derive analytic
expressions for the entanglement entropy $S_{(n)}$ both for $n,L\gg1 $ and for
$n\gg1,L=\infty$. As a result, we find that in the ground state sector with a
fixed value of $S^{z}$ the block entanglement entropy diverges for large $n$
as $S_{(n)}=\sigma\log_{2}[2\pi e n(L-n)/L]+C$. In section 4 the case of spin
$\sigma\geq1$ is treated in more detail. A discussion and some further remarks
close the paper.

\section{Permutation invariant states}

Let us consider states in a Hilbert space
$({\mathchoice{\setbox0=\hbox{$\displaystyle\rm C$}\hbox{\hbox
to0pt{\kern0.4\wd0\vrule height0.9\ht0\hss}\box0}}{\setbox0=\hbox{$\textstyle\rm C$}\hbox{\hbox
to0pt{\kern0.4\wd0\vrule height0.9\ht0\hss}\box0}}%
{\setbox0=\hbox{$\scriptstyle\rm C$}\hbox{\hbox
to0pt{\kern0.4\wd0\vrule height0.9\ht0\hss}\box0}}%
{\setbox0=\hbox{$\scriptscriptstyle\rm C$}\hbox{\hbox
to0pt{\kern0.4\wd0\vrule height0.9\ht0\hss}\box0}}}^{d})^{\otimes L}$ of a
quantum spin chain of local spin $\sigma$, where $d=2\sigma+1$ is the
dimension of the spin space at every site $i$ and $d^{L}$ the dimension of the
whole Hilbert space. Permutation invariant states constitute a subspace $Q$ of
substantially smaller dimension
\begin{equation}
\kappa(L)=\dim Q=\binom{L+2\sigma}{2\sigma}, \label{kappa(L)}%
\end{equation}
this being the number of possible ways to distribute $L$ indistinguishable
objects among $2\sigma+1$ boxes. Denoting by $N_{j}$ the number of objects in
the block $j$, the state of the whole system is completely characterized by
\begin{align}
|\Psi(L,N_{0},N_{1},  &  ...,N_{2\sigma})\rangle=\sqrt{\frac{N_{0}%
!N_{1}!...N_{2\sigma}!}{L!}}\times\label{Psi_permutation}\\
&  \sum_{P}|\underbrace{\downarrow\downarrow...\downarrow}_{N_{0}%
}...\underbrace{\nearrow\nearrow...\nearrow}_{N_{i}}...\underbrace
{\uparrow\uparrow...\uparrow}_{N_{2\sigma}}\rangle,\nonumber
\end{align}
where $N_{0}$ is the number of spins pointing down ($\sigma_{z}=-\sigma$),
$N_{j}$ is the number of spins with $\sigma_{z}=-\sigma+j$ , up to
$N_{2\sigma}$ spins with maximal $\sigma_{z}=\sigma$, occupying in total
$L=N_{0}+N_{1}+...+N_{2\sigma}$ sites. The sum is taken over all possible
permutations, the total number of which is $\frac{L!}{N_{0}!N_{1}%
!...N_{2\sigma}!}$, with the prefactor in (\ref{Psi_permutation}) taking care
of the normalization.

We will be interested in the entanglement of a block of $n$ spins with the
remaining $L-n$ spins (playing here the role of the environment),
characterized by the von Neumann entropy
\begin{equation}
S_{(n)}=-tr(\rho_{n}\log_{2}\rho_{n})=-\sum_{k}\lambda_{k}\log_{2}\lambda_{k},
\label{von-Neumann_entropy}%
\end{equation}
where $\rho_{n}$ is the reduced density matrix of the block, obtained from the
density matrix $\rho$ of the whole system by tracing out the degrees of
freedom of the environment $\rho_{(n)}=tr_{(L-n)}\rho$, and $\lambda_{k}$ are
its eigenvalues. The density matrix of the whole system is a projector on the
pure state in (\ref{Psi_permutation}), i.e. $\rho=|\Psi(L,N_{0},N_{1}%
,...N_{2\sigma})\rangle\langle\Psi(L,N_{0},N_{1},...N_{2\sigma})|$. Notice
that due to the permutational symmetry, $S_{(n)}$ does not depend on the
particular choice of the sites in the block but only on its size $n$. The
eigenvalues $\lambda_{k}$ of the reduced density matrix are by construction
all real, nonnegative, and sum up to one: $\sum_{k}\lambda_{k}=1$.

\bigskip Before giving the general expression for $\lambda_{k}$ we discuss
separately the two-states case of $\sigma=1/2$. In this case the state of the
system (\ref{Psi_permutation}) reduces to
\begin{equation}
|\Psi(L,N)\rangle=\binom{L}{N}^{-1/2}\sum_{P}|\underbrace{\uparrow
\uparrow...\uparrow}_{N}\underbrace{\downarrow\downarrow...\downarrow\rangle
}_{L-N} \label{GS}%
\end{equation}
where $\uparrow$ and $\downarrow$ denote spin up and spin down respectively.
This state appears in the literature in different physical situations. Since
the entanglement properties do not depend on the underlying model but only on
the form of the wavefunction, we list some models for which (\ref{GS}) is an
exact ground state to show the diversity of applications.

\noindent i) The isotropic Heisenberg ferromagnet, $H=J\sum_{i{=1}}%
^{L}\overset{\rightarrow}{\sigma_{i}}\overset{\rightarrow}{\sigma_{i+1}}$
where $\sigma_{i} $ are Pauli matrices, $J<0$ denotes the exchange constant
and $L$ is the total number of spins. (We assume periodic boundary conditions
$L+1\equiv1 $.) The ground state belongs to a multiplet of total spin $S=L/2$
and is $(L+1)$-fold degenerate with $S^{z}=-\frac{L}{2},-\frac{L}%
{2}+1,...\frac{L}{2}$. The state (\ref{GS}) is a pure state, corresponding to
the multiplet component with a fixed number $N$ of spins up.

\noindent ii) The Heisenberg antiferromagnet at $\Delta=-1$. In this case, the
state (\ref{GS}) is obtained from the ground state by a unitary
transformation, inverting every other spin along the chain (note that the von
Neumann entropy is invariant under unitary transformations of the state of the system).

\noindent iii) The generalized Hubbard model in the limit of strong attraction
(the so-called eta-pairing states \cite{Albertini}).

\noindent iv) Hardcore bosons on a complete graph \cite{Toth}, \cite{Penrose}%
,\cite{Mario}.

\noindent v) The Lipkin-Meshkov-Glick model \cite{LMG}.

The state (\ref{GS}) also appears in a classical context as stationary
distribution of the asymmetric exclusion process (ASEP) describing a Markov
process of nonequilibrium stochastic motion of $N$ particles on a ring with
$L$ sites with hardcore exclusion. In this case (\ref{GS}) means that all
particle configurations have equal probabilities in the stationary state
\cite{Gunter_Hamiltonian_approach}. The block entanglement entropy for the
case of spin $\sigma=1/2$ was obtained in \cite{Entanglement_Heisenberg}. In
particular, it was shown that the eigenvalues $\lambda_{k}$ of the reduced
density matrix $\rho_{n}$ are $\lambda_{k}(L,n)=C_{k}^{n}C_{N-k}^{L-n}\left/
C_{N}^{L}\right.  ,\;$ where $C_{m}^{n}$ denotes the binomial coefficient
$n!/(m!(n-m)!)$ and $k=0,1,...\min(n,N)$. In the limit of large $n$,
the von-Neumann entropy was found to be
\begin{equation}
S_{(n)}(p)\approx\frac{1}{2}\log_{2}(pq)+\frac{1}{2}\log_{2}2\pi
e\frac{n(L-n)}{L}.\label{entropy_L}%
\end{equation}
where $p=\frac{N}{L},q=1-p$. In the next section we generalize these results
to the case of arbitrary spins.

\section{Entanglement entropy for arbitrary on-site spin $\sigma$}

We formulate the main result of this section in the following

\noindent\textbf{Theorem}: \textit{The eigenvalues of the reduced density
matrix} $\rho_{n}(N_{0},N_{1},...N_{2\sigma})$ \textit{of a block of}
$n$ \textit{spins in the permutation invariant state
(\ref{Psi_permutation}) of the whole system are given by}
\begin{equation}
\lambda_{\mathbf{k}}(L,n,\mathbf{\sigma})=\frac1 {C^{L}_{n}} {\prod_{i=0}
^{2\sigma}C^{N_{i}}_{k_{i}}} \label{eigenvalues_finite}%
\end{equation}
where $k_{i},N_{i}$ satisfy $k_{0}+...+k_{2\sigma}=n$, $N_{0}+...+N_{2\sigma
}=L$. Here and below we denote the set of $k_{i}$ by the bold $\mathbf{k}$.
To prove the theorem we note that the reduced density matrix $\rho_{n}$ is
decomposed with respect to symmetric orthogonal subspaces of the system of $n$
spins, classified by the numbers $k_{j}=0,1,...\min(n,N_{j})$ of spins with
$\sigma_{z}=-\sigma+j$ in the block
\begin{equation}
\rho_{n}(N)=\sum_{\mathbf{k}}w_{\mathbf{k}}|\psi(n,\mathbf{k})\rangle
\langle\psi(n,\mathbf{k})|. \label{decomposition}%
\end{equation}

\begin{figure}[ptbptb]
\begin{center}
\includegraphics[
height=2.5414in, width=3.0566in ] {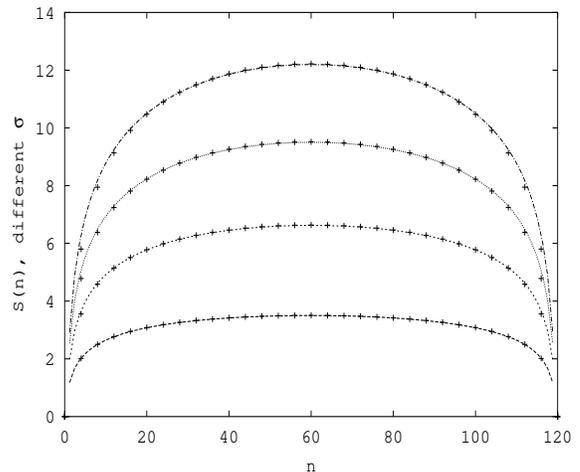}
\end{center}
\caption{Entanglement entropy as function of a number of sites $n$ in the
block, for different on-site spin $\sigma=\frac{1}{2},1,\frac{3}{2},2$, for
finite system $L=120$. Comparison of the exact formula (points) with the
limiting expression (\ref{entropy_limit_general}) (continuous curves). In all
cases equal partitions of particles is taken $p_{i}=N_{i}/L=1/(2\sigma+1)$.}%
\label{fig_SfiniteSall}%
\end{figure}

Here the state $|\psi(n,\mathbf{k})\rangle$ has the same structure as
(\ref{Psi_permutation}),
\begin{equation}
|\psi(n,\mathbf{k})\rangle=\sum_{P}|\underbrace{\downarrow\downarrow
...\downarrow}_{k_{0}}...\underbrace{\nearrow\nearrow...\nearrow}_{k_{i}
}...\underbrace{\uparrow\uparrow...\uparrow}_{k_{2\sigma}}\rangle
\end{equation}
and $w_{k}$ is the corresponding probability, given by the number of ways one
can distribute $N_{j}$ spins (of values $-\sigma+j$) on $k_{j}$ sites for all
possible values of $N_{j},k_{j}$, $j=0,...,2\sigma$, divided by the number of
ways one can distribute $N=N_{0}+...+N_{2\sigma}$ spins on $n=k_{0}%
+...+k_{2\sigma}$ sites (the total number of states), i.e.
\begin{equation}
w_{\mathbf{k}}=\frac{1}{C^{L}_{n}}{\prod_{j=0}^{2\sigma}C^{N_{j}}_{k_{j}}}.
\end{equation}
Since the states $|\psi(n,\mathbf{k})\rangle$ are orthogonal, the representation in
(\ref{decomposition}) is diagonal and the $w_{\mathbf{k}}$ coincide with the
eigenvalues of $\rho_{n}(N)$.  This proves of the theorem. Notice
that for the case $\sigma=1/2$ Eq. (\ref{eigenvalues_finite}) reproduces the
results derived in \cite{Entanglement_Heisenberg} (use $N_{0}=L-N_{1}$,
$k_{0}=n-k_{1}$ and the invariance under the exchange of $N_{1}$ and $n$).
\begin{figure}[ptb]
\begin{center}
\includegraphics[
height=2.5414in, width=3.0566in] {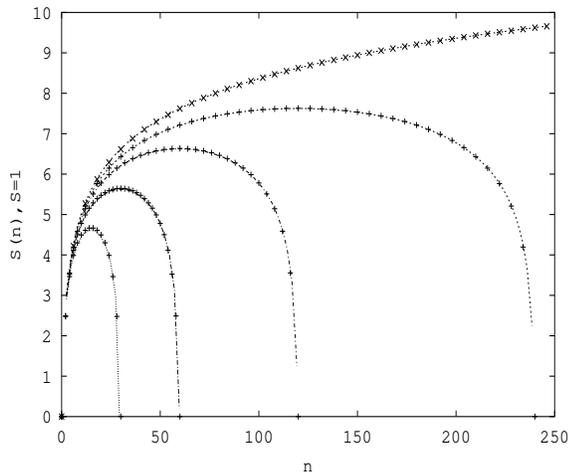}
\end{center}
\caption{Exact (points) versus analytical (curves) comparison of von-Neumann
entropy $S_{(n)}$ for spin $\sigma=1$ in systems with $L=30,60,120,240,\infty
$. Blocks of all sizes $0\leq n\leq L$ (for finite $L$) are considered. The
other parameters $\frac{N_{1}}{L}=\frac{N_{2}}{L}=\frac{1}{3}$. }%
\label{fig_sfiniteS1}%
\end{figure}Having found the eigenvalues of $\rho_{n}(N)$ one can compute the
entanglement entropy $S_{(n)}$ for arbitrary $L,n$ and $N$ . In the large-$n$
limit one obtains, see the Appendix, in analogy with the case of $\sigma=1/2$
\cite{Entanglement_Heisenberg}, that
\begin{align}
S_{(n)}(L,\mathbf{\sigma})  &  =C+\sigma\log_{2}\left(  2\pi e\frac{n(L-n)}%
{L}\right)  ,\text{ \ \ \ \ \ }n\gg1\text{,}\label{entropy_limit_general}\\
C  &  =\frac{1}{2}\log_{2}\left(  {\displaystyle\prod\limits_{k=0}^{2\sigma}%
}p_{k}\right)  , \label{C(N)}%
\end{align}
where $p_{k}=\frac{N_{k}}{L}$. In Figs. \ref{fig_sfiniteS1}
,\ref{fig_SfiniteSall} we compare the exact entropy of finite systems, as
computed from the exact expressions Eqs. (\ref{von-Neumann_entropy},
\ref{eigenvalues_finite}), with the analytical expression
(\ref{entropy_limit_general}), from which we see that there is an excellent
agreement also for small values of $n{\displaystyle\prod\limits_{k=0}%
^{2\sigma}}p_{k}$. In the thermodynamic limit $L\rightarrow\infty,\frac{N_{i}%
}{L}\rightarrow p_{i}$, the eigenvalues of the reduced density matrix
(\ref{eigenvalues_finite}) simplify to
\[
\lambda_{\mathbf{k}}(\infty,n,\{p_{i}\}_{i=0}^{2\sigma})=n!{\displaystyle\prod
\limits_{i=0}^{2\sigma}}\frac{(p_{i})^{k_{i}}}{k_{i}!}%
\]
The entanglement entropy then becomes
\begin{equation}
S_{(n)}(\infty,\{p_{i}\}_{i=0}^{2\sigma})=C(\{p_{i}\})+\sigma\log_{2}\left(
2\pi en\right)   \label{Entropy_L_Infinite}%
\end{equation}
where the constant $C$ is given by (\ref{C(N)}). 

Notice that while the general formulae (\ref{von-Neumann_entropy}%
,\ref{eigenvalues_finite}) are valid for an arbitrary choice of parameters, 
the analytical result (\ref{entropy_limit_general}) is valid for $n%
{\displaystyle\prod\limits_{k=0}^{2\sigma}}
p_{k}\gg1$. If in the limit of an infinite system $L\rightarrow\infty$ some of
the $p_{k}$ vanish (say, $z$ coefficients $p_{k}\rightarrow0$ out of a total
number of coefficients $2\sigma$), the entangled state of $n$ spins will
effectively behave like the one with effective site spin $\tilde{\sigma
}=\sigma-\frac{z}{2}$, and the von Neumann entropy will respectively grow as
$S_{n} \sim(\sigma-\frac{z}{2})\log_{2}n$ instead of $S_{n}\sim\sigma\log
_{2}n$, see (\ref{Entropy_L_Infinite}).

It is instructive to recall that Eq. (\ref{entropy_limit_general}) is derived
under the assumption of a \textit{pure }state of the global system. If the
state of the whole system is mixed in the ensemble of states with the same
symmetry (\ref{Psi_permutation}), the resulting von-Neumann entropy $S_{(n)}$
will become larger. An upper bound of $S_{(n)}$ can be derived by noting that
the number of nonzero eigenvalues of the reduced density matrix is equal to
the number of terms in the decomposition (\ref{decomposition}), which is
$\kappa(n)$, see (\ref{kappa(L)}). The upper limit of the entanglement entropy
is thus reached when all $\lambda_{k}$ in (\ref{von-Neumann_entropy}) are
equal, implying
\begin{equation}
\sup(S_{(n)})=\log_{2}\binom{n+2\sigma}{2\sigma}.\label{maxentropy}%
\end{equation}
This limit is attained for a mixed state of the global system with all
components of the ensemble (\ref{Psi_permutation}) equally weighted. In this
case the von-Neumann entropy is given by the above expression even for finite
$n,L$ (see \cite{Entanglement_Heisenberg} for an example). For large $n$,
$\sup(S_{(n)})\approx2\sigma\log_{2}n$.

We also remark that in the critical spin models where spin-spin correlation
decay algebraically in the ground state (e.g., the region $-1<\Delta\leq1$ of 
the
antiferromagnetic Heisenberg chain), there are three distinct physical
properties contributing to the entanglement: (a) on-site correlations due to
single-site quantum fluctuations, (b) algebraically decaying spin-spin
correlations which survive in the thermodynamic limit and (c) the correlations
due to the constraint of fixed magnetization which vanish in every domain of
finite size in the thermodynamic limit. For permutation invariant states
(\ref{Psi_permutation}) only contribution (a) is left in the thermodynamic
limit for finite domains, but for domains of finite volume fraction also the
correlations (c) due to the constraint of fixed magnetization remain relevant.

\section{Models with higher spin}

As remarked above the entanglement entropy of a quantum state does not depend
on any underlying model but only on the properties of that state. Nevertheless
it is of interest to have some insight for which systems the permutation
invariant states (\ref{Psi_permutation}) considered here are the ground state
of that quantum system. First of all, the generalizations of the
Sutherland model
\cite{Sutherland} describing quantum spin chains with an interaction given by
the permutation operator in $SU(N)$,
\begin{equation}
H=\sum J_{ij}P_{i,j} \label{H_Sutherland}%
\end{equation}
are obviously invariant under the permutation group $S_{N}$. Here
the set of $J_{ki}$ is defined on any connected graph, an example being
nearest neighbor interaction.
For ferromagnetic
interaction, all $J_{ij}<0$, the states (\ref{Psi_permutation}) span its
ground state, which can be proved along the lines of \cite{Pratt}. For $SU(2)$
and $J_{ij}=J\delta_{i,j+1}$ the Hamiltonian (\ref{H_Sutherland}) reduces to
the isotropic Heisenberg Hamiltonian. For the general case we recall that the
permutation operator is written in terms of spin operators as
\[
P=\sum_{i=0}^{2S}(-1)^{2S+i}{\displaystyle\prod\limits_{k\neq i}^{2S}}%
\frac{2\left(  \mathbf{\vec{S}}\otimes\mathbf{\vec{S}}\right)
-k(k+1)+2S(S+1)}{i(i+1)-k(k+1)}%
\]
where $N=2S+1$, (see, e.g., \cite{BatchelorMaslen}).
For the above example, all states (\ref{Psi_permutation}) are eigenstates of
(\ref{H_Sutherland}) with the lowest energy. 

The increase of entanglement entropy compared to the $s=1/2$ case appears to
result from the larger local state space of the $SU(N)$ chain, but not from
the $SU(N)$-symmetry itself. This picture is supported by the generalized
disordered SU(2)-symmetric spin-$\sigma$ Heisenberg ferromagnet
\begin{equation}
H = \sum_{i,j} J_{i,j} g(\overset{\rightarrow}{\sigma_{i}}\overset
{\rightarrow}{\sigma_{j}}-1)
\end{equation}
where the exchange energies $J_{i,j}\leq0$ may be non-zero between
\textit{any} pair of lattice sites on an arbitrary lattice and $\overset
{\rightarrow}{\sigma_{i}}$ are local $SU(2)$ generators in the spin-$\sigma$
representation. For any polynomial function $g$ with positive expansion
coefficients linear combinations of the permutation invariant states
(\ref{Psi_permutation}) with fixed total magnetization are ground states of
the Hamiltonian, see \cite{Schu94} for a detailed discussion of the ground
states of this model in a probabilistic setting. Since these are not pure
states in the sense discussed above (with all quantum numbers $N_{k}$ fixed)
the entanglement entropy of these ground states is higher than those of the
pure ground states of the $SU(N)$ spin chain.

Other models with pair interaction, but no symmetry: Using the
Perron-Frobenius theorem it is straightforward to construct quantum
Hamiltonians of the structure
\begin{equation}
H=\sum_{i,j}J_{i,j}g_{i,j}%
\end{equation}
where $g_{i,j}$ is a hermitian pair interaction matrix satisfying
$g_{i,j}|\;s\;\rangle=0$ for all $i,j$ and where the wave
function $|\;s\;\rangle$ which is constant for all spin configurations in
$({\mathchoice {\setbox0=\hbox{$\displaystyle\rm C$}\hbox{\hbox
to0pt{\kern0.4\wd0\vrule height0.9\ht0\hss}\box0}} {\setbox0=\hbox{$\textstyle\rm C$}\hbox{\hbox
to0pt{\kern0.4\wd0\vrule height0.9\ht0\hss}\box0}} {\setbox0=\hbox{$\scriptstyle\rm C$}\hbox{\hbox
to0pt{\kern0.4\wd0\vrule height0.9\ht0\hss}\box0}} {\setbox0=\hbox{$\scriptscriptstyle\rm C$}\hbox{\hbox
to0pt{\kern0.4\wd0\vrule height0.9\ht0\hss}\box0}}}^{d})^{\otimes L}$ 
is the ground state of $g_{i,j}$. Then, if all $J_{i,j}\leq0$, the vector
$|\;s\;\rangle$ is also the ground state of $H$, and if furthermore $g_{i,j}$
has no invariant subspaces, it is the unique ground state with maximal
entanglement entropy (\ref{maxentropy}). Such quantum systems also have a
probabilistic interpretation as generator of some irreducible Markov chain
\cite{Gunter_Hamiltonian_approach}.

\section{Summary and conclusions}

We have obtained exact eigenvalues of the reduced density matrix for
permutation invariant states, for arbitrary length of the system. In the
thermodynamic limit, it was shown that the von Neumann entropy of entanglement
of a block of $n$ spins $\sigma$ with the environment grows logarithmically
fast $S_{(n)}\sim\gamma\log_{2}n$ , with a prefactor $\gamma=\sigma$ for a
pure global state and with $\gamma=2\sigma$ for homogenously (maximally) mixed
global state. Various models, the ground states of which are permutation
invariant, are given, for spin $\sigma=1/2$ and higher. We note that the
logarithmic growth of entanglement entropy due to permutation invariance is
\textit{faster} than the one of critical (conformally invariant) models, where
$S_{(n)}\sim\gamma\log_{2}n$ with $\gamma=1/3$ was observed \cite{Vidal}.

It is also interesting to compare the finite size corrections of the entropy
of the permutation invariant states (\ref{entropy_limit_general}) with those
of critical spin chains, $S_{(n)}^{cr}\sim\frac{c}{3}\log_{2}(\frac{L}{\pi
}\sin(\frac{\pi n}{L}))$, obtained in \cite{Cardy} , see Eq.(3.8) in this
paper. Expanding in the first nonvanishing order of $1/L$ both expressions, we
get the finite size corrections $\Delta^{cr}(n)=S_{(n)}^{cr}(L)-S_{(n)}%
^{cr}(\infty)$, \ $\Delta^{per}(n)=S_{(n)}^{per}(L)-S_{(n)}^{per}(\infty)$ for
permutation invariant and critical models as
\begin{align*}
\Delta^{per}(n)  &  =\sigma\log_{2}(1-\frac{n}{L})\approx-\sigma\frac{1}{\ln
2}\frac{n}{L}+O\left(  \frac{n}{L}\right)  ^{2}\\
\Delta^{cr}(n)  &  \approx\frac{c}{3}\log_{2}(1-\frac{1}{3}(\frac{\pi n}%
{L})^{2})\approx-\frac{c}{9}\frac{1}{\ln2}\left(  \frac{\pi n}{L}\right)
^{2}+O(\frac{n}{L})^{4}%
\end{align*}
Thus, besides the difference of the coefficient of the principal logarithmic
divergence ($\gamma\log_{2}n$ with $\gamma=1/3$ for critical $XXZ$ and
$\gamma=\sigma$ for permutation invariant states resp., a substantial
difference in the finite size corrections (linear in $n$ and of order $1/L$
for permutation invariance) and quadratic in $n$ and of order $1/L^{2}$ resp.
(for conformal invariant critical states) is also observed.

\vskip.5cm \acknowledgments \noindent VP acknowledges the INFM, Unit\'{a} di
Salerno, for providing a three months grant during which this work was
initiated, and the Department of Physics of the University of Salerno for the
hospitality. MS acknowledges partial financial support from Forschungszentrum
J\"{u}lich, from the University of Cologne within SFB/TR12 project "Symmetries
and Universality in Mesoscopic Systems" and from a MURST-PRIN-2003 Initiative.

\section{Appendix}

To prove Eq. (\ref{entropy_limit_general}) we first give the general scheme
and then demonstrate the result on a particular example. For simplicity we
shall discuss the limit when the size of the global system tends to infinity,
$L\rightarrow\infty$. The eigenvalues (\ref{eigenvalues_finite})
$\lambda_{\mathbf{k}}\equiv\lambda(k_{0} ,k_{1},...k_{2\sigma})$ for large
values of $n$ and $k_{i}$, can be approximated by a multi-dimensional Gaussian
distribution with mean and moments given by
\begin{equation}
\langle k_{i}\rangle=np_{i} \label{kmean}%
\end{equation}
\begin{equation}
\langle\left(  k_{i}-\langle k_{i}\rangle\right)  ^{2}\rangle=np_{i}\left(
1-p_{i}\right)  \label{kdispersion}%
\end{equation}
\begin{equation}
\langle\left(  k_{i}-\langle k_{i}\rangle\right)  \left(  k_{j}-\langle
k_{j}\rangle\right)  \rangle=-np_{i}p_{j}\text{, \ }i\neq j
\label{kdispersion_nondiagonal}%
\end{equation}
where $p_{i}=\frac{N_{i}}{L}$, $i=0,1,...2 \sigma$, is the average density of
spins $-\sigma+i$ in the system. To prove Eqs. (\ref{kmean}%
)-(\ref{kdispersion_nondiagonal}) we first observe that, using the Stirling
approximation $m!=m^{m} exp(-m) \sqrt{2 \pi m}$, in the limit $L, N_{i}
\rightarrow\infty$ the eigenvalues (\ref{eigenvalues_finite}) of the reduced
density matrix can be approximated as
\begin{equation}
\lambda_{\mathbf{k}}\approx c_{\mathbf{k}}(n) \prod_{i=0}^{2 \sigma}
p_{i}^{k_{i}}%
\end{equation}
with $c_{\mathbf{k}}(n)= n!/(k_{0}! k_{1}! ... k_{2 \sigma}!)$. Notice that in
this limit the distribution of the $\lambda_{\mathbf{k}}$ coincides with the
multinomial distribution
\begin{align}
\left(  \sum_{i=0}^{2\sigma} p_{i}\right)  ^{n}  &  =\sum_{\mathbf{k}}
c_{\mathbf{k}}(n) \prod_{i=0}^{2 \sigma} p_{i}^{k_{i}}= \sum_{\mathbf{k}}
\lambda_{\mathbf{k}}=1 .\nonumber
\end{align}
Using this expression, the mean value of $k_{\alpha}$, $\alpha=0,1,...,2
\sigma$, can be found as
\begin{align}
\langle k_{\alpha}\rangle=\sum_{\mathbf{k}} k_{\alpha} c_{\mathbf{k}}(n)
\prod_{i=0}^{2 \sigma} p_{i}^{k_{i}}= p_{\alpha}\frac{\partial}{\partial
p_{\alpha}} (\sum_{i=0}^{2\sigma} p_{i}) ^{n} =n p_{\alpha}.\nonumber
\end{align}
In a similar manner we obtain
\begin{align}
&  \langle k_{\alpha}\left(  k_{\alpha}-1\right)  \rangle=p_{\alpha}^{2}
\frac{\partial^{2}}{\partial p_{\alpha}^{2}} \left(  \sum_{i=0}^{2\sigma}
p_{i} \right)  ^{n}=n(n-1) p_{\alpha}^{2},\nonumber\\
&  \langle k_{\alpha}k_{\beta}\rangle=p_{\alpha} p_{\beta}\frac{\partial^{2}%
}{\partial p_{\alpha}\partial p_{\beta}} \left(  \sum_{i=0}^{2\sigma} p_{i}
\right)  ^{n}= n(n-1)p_{\alpha}p_{\beta},\nonumber
\end{align}
from which Eqs. (\ref{kdispersion}), (\ref{kdispersion_nondiagonal}) readily
follow. Denoting by $x_{i}=k_{i}/n$, so that $0<x_{i}<1$, the eigenvalues
$\lambda_{\mathbf{k}}$ \ are approximated as
\begin{equation}
\lambda_{\mathbf{k}}\approx\frac{\sqrt{\det A}}{(2\pi)^{\sigma}}\exp\left(
-\frac{1}{2} {\displaystyle\sum\limits_{i,j=1}^{2\sigma}} A_{ij}\left(
x_{i}-p_{i}\right)  \left(  x_{j}-p_{j}\right)  \right)  \label{Gaussian}%
\end{equation}
where the shifting of $x_{i},x_{j}$ is introduced to account for the nonzero
mean (\ref{kmean}), and the coefficients of the symmetric matrix
$a_{ij}=a_{ji}$ are fixed using the moments (\ref{kdispersion}),
(\ref{kdispersion_nondiagonal}). On the other hand, the computation of the
moments from the distribution (\ref{Gaussian}) gives
\begin{equation}
\langle\left(  x_{i}-p_{i}\right)  \left(  x_{j}-p_{j}\right)  \rangle
=\frac{M_{ij}}{\det A} \label{Minors}%
\end{equation}
where $M_{ij}$ are the minors of the matrix $A$. Comparing Eq. (\ref{Minors})
with Eqs. (\ref{kdispersion} ),(\ref{kdispersion_nondiagonal}), we can fix the
elements $A_{ij}$ of the matrix. To illustrate the method let us consider the
case of spin $\sigma=1$. Then,
\[
\lambda_{\mathbf{k}}\equiv\lambda(x,y)\approx\frac{\sqrt{D}}{2\pi}\exp\left(
-\frac{1}{2}\left(  ax^{2}+2bxy+cy^{2}\right)  \right)
\]
where we introduced $x=x_{1}-p_{1},$ $y=x_{2}-p_{2}$ and $D=\det A=ac-b^{2}$
for brevity of notation. From (\ref{kdispersion}%
),(\ref{kdispersion_nondiagonal}) we have $\langle x^{2}\rangle=np_{1}%
(1-p_{1})$, $\langle y^{2}\rangle=np_{2}(1-p_{2})$, and $\langle
xy\rangle=-np_{1}p_{2}$. On the other hand, from (\ref{Minors}) we obtain
$\langle x^{2}\rangle=c/D$, $\langle y^{2}\rangle=a/D$, and $\langle
xy\rangle=-b/D$. Computing the determinant%
\[
D=ac-b^{2}=D^{2}\left(  \langle x^{2}\rangle\langle y^{2}\rangle-\langle
xy\rangle^{2}\right)
\]
and substituting the moments, we get
\[
D^{-1}=n^{2}p_{1}p_{2}p_{3}\text{.}%
\]
The von Neumann entropy is then computed as
\[
S\approx-\int dx\int dy\left(  \lambda(x,y)\log_{2}\lambda(x,y)\right)
\]
Notice that for finite $p_{i}$ the larger contribution to the integral comes
from the neighbor of the origin $x=y=0$ so that we can extend the limits of
integration to the whole real axis and perform the integral exactly. This
leads to
\[
S(n)\approx\log_{2}\frac{2n\pi e}{\sqrt{D}}=\log_{2}2\pi ne+\frac{1}{2}%
\log_{2}\left(  p_{1}p_{2}p_{3}\right)  ,
\]
which coincides with the expression (\ref{entropy_limit_general}) in the limit
$L=\infty$ for the case of spin $\sigma=1$. A more detailed analysis,
analogous to the one done in \cite{Entanglement_Heisenberg}, restores the
finite-size dependence on $L$ of the von-Neumann entropy as
\[
S(n)\approx\log_{2}\left(  2\pi e\frac{n(L-n)}{L}\right)  +\frac{1}{2}\log
_{2}\left(  p_{1}p_{2}p_{3}\right)  \text{ for }\sigma=1
\]
Working out the same procedure for an arbitrary spin $\sigma$ we obtain
(\ref{entropy_limit_general}).

\end{document}